\documentclass[draft]{agujournal2019}
\usepackage{url} 
\usepackage{lineno}
\usepackage[finalnew]{trackchanges} 
\usepackage{soul}
\usepackage{amsmath}
\usepackage{amssymb}
\usepackage{xcolor}
\usepackage{tikz}

\newcommand{\corrb}[1]{\textcolor{black}{#1}} 
\draftfalse

\journalname{Geophysical Research Letters}

\begin{document}

%
%

\title{When barchan dunes move over craters}

\textcolor{blue}{An edited version of this paper was published by AGU.\\
	Pl\'acido, P.V.R., Borges, D.S., Assis, W.R., Franklin, E.M., When barchan dunes move over craters. Geophysical Research Letters, 52, e2025GL120187, 2025. . Published under the CC BY-NC-ND 4.0 license, DOI 10.1029/2025GL120187.\\
	To view open-access paper, go to https://doi.org/10.1029/2025GL120187.}

%
%




\authors{P. V. R. Pl\'acido\affil{1}, D. S. Borges\affil{3}, W. R. Assis\affil{2}, E. M. Franklin\affil{1}}


\affiliation{1}{Faculdade de Engenharia Mec\^anica, Universidade Estadual de Campinas (UNICAMP),\\
	Rua Mendeleyev, 200, Campinas, SP, Brazil}

\affiliation{2}{Saint Anthony Falls Laboratory, University of Minnesota,\\
	2 3rd Ave SE, Minneapolis, Minnesota, USA}

\affiliation{3}{Faculty of Physics, University of Duisburg-Essen, \\
	47057 Duisburg, Germany}




\correspondingauthor{Danilo S. Borges}{danilo.dasilvaborges@uni-due.de}




\begin{keypoints}
\item We show that subaqueous barchans can be blocked, destroyed, or pass over craters, with transitional situations
\item When the barchan size is moderate with respect to the crater diameter, flow-induced instabilities break the bedform into smaller dunes
\item Based on a crater-dune size ratio and a modified Stokes number, we propose a map that classifies the different outcomes of interactions
\end{keypoints}

%
%

%
%


\begin{abstract}
We investigate the possible outcomes of a subaqueous barchan moving over a crater-like depression in the bed. For that, we carried out experiments where we varied the dune size, crater and grain diameters, and flow velocities. We found that subaqueous barchans can be blocked, destroyed, or pass over craters, with transitional situations, and that strong instabilities take place under some conditions. Based on a dune-crater size ratio and a modified Stokes number, we propose a map that classifies the different outcomes of interactions. If used with caution, the map can serve as a reference for understanding the much slower behavior of dunes migrating over or near craters on the surface of Mars.
\end{abstract}

\section*{Plain Language Summary}
A great portion of Mars' surface is covered with rocky and sandy landscapes, including craters and migrating dunes, among which we find crescent-shaped dunes known as barchans. Those dunes are shaped by a roughly unidirectional wind over sand, and migrate relatively slowly on Mars (up to about one meter per year). In many instances, barchans are close to or inside craters, and we expect that their morphology and migration rate are affected by the crater. However, due to the large timescales involved, the outcomes of such dune-crater interactions are not known. Taking advantage of the much smaller and faster scales of subaqueous barchans (centimeters and minutes), we conducted experiments in a water channel to understand how a barchan behaves in the presence of a crater. We varied the sizes of dune and crater, grain diameter, and water velocities, and recorded the barchan evolution. We found that the dune can be blocked, destroyed, or pass over the crater, with transitional situations, and we propose a classification map in which the barchan behavior depends on two dimensionless parameters. Our results bring new insights into how Martian dunes interact with impact craters, with implications for sand transport pathways and trapping over geological timescales.

\section{Introduction}

Barchan dunes emerge under unidirectional flows and limited sediment supply, occurring both in eolian and liquid environments and being found on Earth, Mars and other celestial bodies \cite{Bagnold_1, Herrmann_Sauermann, Hersen_3, Elbelrhiti, Claudin_Andreotti, Parteli2, Courrech, Courrech2}. In all cases, barchans have roughly the same crescent shape, which is a strong attractor \cite{Yang_3}, but the scales are considerably different: centimeters and minutes under water, tens of meters and years on Earth's deserts, and up to one kilometer and millenniums on Mars \cite{Hersen_1, Claudin_Andreotti}. In the latter case, because the crescent shape has horns (tips) that point in the downstream direction, barchans have been used as a marker of the wind direction in different regions of Mars \cite{Tsoar, Hayward, Rubanenko, Rubanenko2}.

Those dunes are usually found in barchan fields \cite{Norris, Gay, Vermeesch, Hugenholtz}, where barchan-barchan interactions regulate their size \cite{Endo2, Hersen_2, Hersen_5, Kocurek, Genois, Genois2, Assis, Assis2, Assis3, Marvin, Assis5, Assis6}. Moreover, barchan-barchan interactions are responsible for variations in the dune morphology, with different outcomes depending on the size ratio of dunes and flow conditions, creating longer, wider, and/or asymmetric barchans \cite{Assis, Assis2}. Besides dune-dune interactions, barchans can reach dune-size obstacles, such as bridge pillars in subaqueous environment \cite{Rubi, Jia}, buildings on Earth \cite{Raffaele}, and rocks and crater rims on Mars \cite{Breed, Urso, Roback}. Recently, \citeA{Assis4} investigated experimentally the interaction of a subaqueous barchan with an obstacle, for different sizes and shapes, flow velocities, and grain types, and found that barchans are blocked, bypass, or pass over the obstacle depending on the grain type, fluid flow, and relative sizes of barchans and obstacles. They also proposed an \textit{ad hoc} classification map based on a modified Stokes number and an obstacle-dune size ratio. More recently, \citeA{Assis6} carried out grain-scale simulations and showed that a strong vortex resulting from the interaction between the recirculation region downstream the dune and a horseshoe vortex upstream the obstacle makes the grains circumvent the obstacle in the bypass and trapping (blocked) cases.

Despite the recent findings on barchan-barchan and barchan-obstacle interactions, another scenario remains unexplored in controlled experiments: the behavior of barchan dunes migrating near and over crater-like depressions. Such configurations are present in the Martian landscape, where dunes are found close to or inside craters \cite{Fenton, Chojnacki2, Chojnacki, Cardinale, Roback, Gunn, Rubanenko2, Love}. In addition to their elevated rim, craters have a concave geometry that significantly disturbs the fluid flow and sediment transport \cite{Gundersen, Rubanenko2}, potentially inducing unique morphodynamic responses, including partial collapse, circumvention, and trapping in recirculation zones. We note that the disturbances caused by crater-like depressions are different from those due to obstacles placed above the ground investigated recently. In this letter, we report an experimental investigation of the outcomes of a subaqueous barchan moving over a crater-like depression in the bed. The experiments were carried out in a water tunnel, and we varied the dune size, crater and grain diameters, and flow velocities. We show that, depending on the experimental conditions, subaqueous barchans can be blocked, destroyed, or pass over craters, with transitional situations. For the latter, we find that strong instabilities take place, in which the bedform is split into smaller bedforms. Finally, we find the characteristic time for interactions, and propose a map based on a dune-crater size ratio and a modified Stokes number that classifies the different outcomes of the barchan-crater interactions. Although obtained for water, extrapolations of the classification map can serve as a reference for understanding the behavior of barchans migrating over craters on Mars, shedding new light on the problem.

\section{Materials and Methods}

Experiments were performed in a recirculating water channel, in which water is pumped from a water reservoir into a 5-m-long closed-conduit channel, and then returns to the water tank \cite<the same experimental setup used in>{Assis, Assis4}. The channel is of transparent material, with a rectangular cross section (width = 160 mm and height 2$\delta$ = 50 mm), and its last 2 m consist of the 1-m-long test section followed by a 1-m-long discharging section. Therefore, the test section begins at approximately 40 hydraulic diameters (40 $\times$ 3.05$\delta$) from the channel entrance, assuring a fully-developed turbulent flow at the entrance of the test section. The test section can be removed and replaced, so that we manufactured three different test sections, each one with a different crater machined on the bottom wall (centered in the channel centerline). With a given test section in place, the channel was filled with water and, for each test, glass spheres were poured inside, forming a granular pile upstream the crater (also centered in the channel centerline, at a distance of approximately its own size from the crater). By turning the pumps on, a turbulent water flow was imposed, deforming the initial pile into a barchan dune that migrated toward the crater. Images of the bedforms were recorded by a camera placed above the channel, and their morphology, position and velocity were measured by image processing. A layout of the experimental setup, a photograph of the test section, and microscopy images of the used grains are shown in Figures S1 to S4 in the Supporting Information.

In the experiments, the water was within 21 and 34 $^o$C, and we used glass spheres (density $\rho_s$ = 2500 kg/m$^3$) with diameters within $0.40$ mm $\leq\,d\,\leq$ $0.60$ mm and within 0.15 mm $\leq\,d\,\leq$ 0.25 mm, separated in two different populations. The initial mass $m$ of dunes varied between 1 and 40 g, corresponding to piles roughly conical, with equivalent basal diameter $D$ within 14 and 48 mm (computed based on the total mass poured), while the diameters $D_c$ of the circular craters were either 50mm or 100 mm, and the rectangular crater had a width $W_c$ $=$ 160 mm and length $L_c$ $=$ 300 mm, corresponding to an equivalent diameter $D_c$ $=$ 300 mm (since $W_c$ $>>$ $D$). The base of craters were plane, at a depth $H_c$ $=$ 5 mm from the bottom of the channel, and the crater walls had a 30$^\circ$ slope (drawings of the craters are available in Figures S6 to S8 in the Supporting Information). Although simplified, these craters have a geometry similar to some of those found in nature \cite{Arvidson, Melosh}. The bulk velocities of water (cross-sectional mean velocities) $U$ varied within 0.191 m/s and 0.347 m/s, corresponding to Reynolds numbers based on the channel height, Re $=$ $\rho U D_h /\mu$, within 1.5 $\times$ 10$^4$ and 3.2 $\times$ 10$^4$, respectively, where $D_h$ is the channel hydraulic diameter \cite{Schlichting_1} and $\rho$ and $\mu$ Pa$\cdot$s are the density and dynamic viscosity of water (for which we considered the temperature variations). The shear velocities on the channel walls $u_*$ were shown in previous works \cite{Franklin_9, Cunez2, Alvarez3} to follow the Blasius correlation \cite{Schlichting_1}, so that the flow was in hydraulic-smooth regime, and found to be in the range 0.0112 m/s $\leq$ $u_*$ $\leq$ 0.0193 m/s. This corresponds to Reynolds numbers at the grain scale, Re$_*$ $=$ $\rho u_* d / \mu$, within 3 and 10, to Shields numbers, $\theta$ = $(\rho u_*^2)/((\rho_s - \rho )gd)$, within 0.017 and 0.127, and to thicknesses of the viscous sublayer, $5l_v$, within 0.22 and 0.44 mm, where $\nu$ $=$ $\mu/\rho$ is the kinematic viscosity and $l_v$ $=$ $\nu/u_*$ is the viscous length. The Stokes numbers \cite{Andreotti_6} based on the bulk velocity, St $=$ $U d \rho_s / (18\mu)$, varied within 9 and 26, the Froude number based on the channel height, Fr $=$ $U/\sqrt{g(2\delta)}$, varied within 0.27 and 0.50, and the saturation length $L_{sat}$ $\approx$ 4.4$(\rho_s/\rho)d$ \cite{Claudin_Andreotti} varied within 2.2 and 5.5 mm. A summary of the pertinent parameters and their ranges is available in Table S1 of the Supporting Information.

Although our experiments were carried out in a closed-conduit channel, we note that most of subaqueous bedforms measured can be classified as ripples \cite{Lapotre}, since, with the exception of the largest barchans, they would not interact with an existing free surface \cite<and bedforms in closed-conduit channels have similar properties as those formed in equivalent open-channel flows,>{Coleman_1}. However, subaqueous bedforms are shaped directly by the action of the fluid flow, as happens also for dunes in eolian environment. In particular, subaqueous barchans share similarities with eolian barchans found in nature, with a well defined crescent shape, same values of aspect ratios, presence of horns, etc. \cite{Hersen_1, Franklin_8, Alvarez}. Therefore, we will use in the following the word \textit{dune} for the barchans and \textit{bedform} for the remaining bedforms.

\section{Results and Discussion}

\begin{figure}[ht]
	\begin{center}
		\includegraphics[width=1\linewidth]{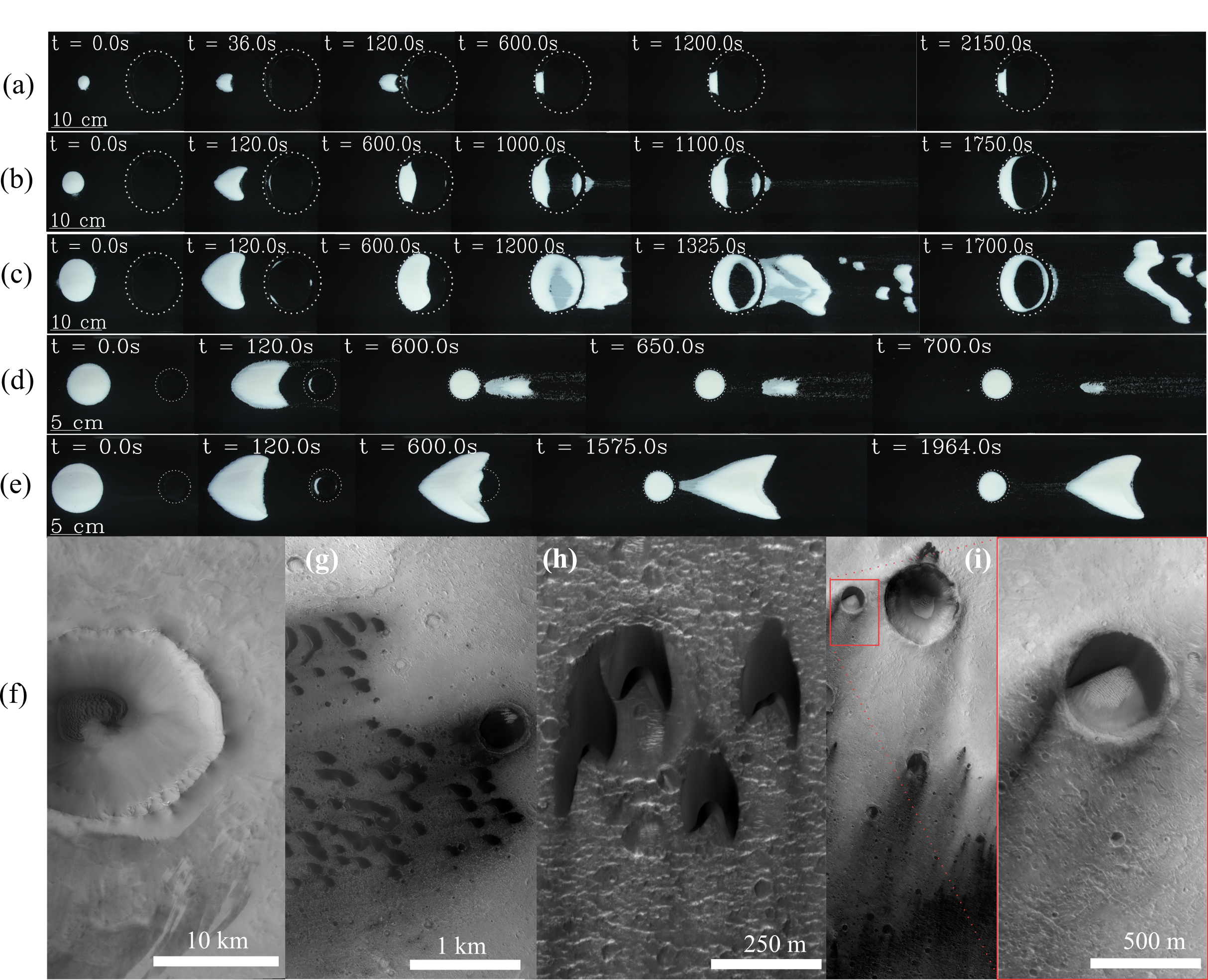}\\
	\end{center}
	\caption{(a)-(e) Snapshots placed side by side showing the time evolution for the Trapped, Trapped with mass loss, Flow-induced fragmentation, Overpassing with mass loss, and Overpassing cases, respectively. The crater rim is highlighted by a dashed line in the snapshots, and the flow is from left to right. (f)-(i) HiRISE image (High Resolution Imaging Science Experiment, https://www.uahirise.org/[Dataset], n.d.) showing examples of barchan dunes interacting with impact craters on the surface of Mars. Morphological patterns in these Martian examples exhibit striking similarities with the regimes shown in (a)-(e). The right side of Panel (i) is a zoom in the region marked (red square) on the left side, and shows a bedform trapped inside a crater. Coordinates: latitudes [71.8741°N, 10.573°N, 23.190°N, 12.884°N], longitudes [345.094°E, 357.246°E, 339.585°E, 356.280°E], spacecraft altitudes [276.0, 287.3, 287.3, 279.5 km]. Courtesy NASA/JPL-Caltech/UArizona.}
	\label{fig:snapshots_craters}
\end{figure}

\begin{figure}[ht]
	\begin{center}
		\includegraphics[width=1\linewidth]{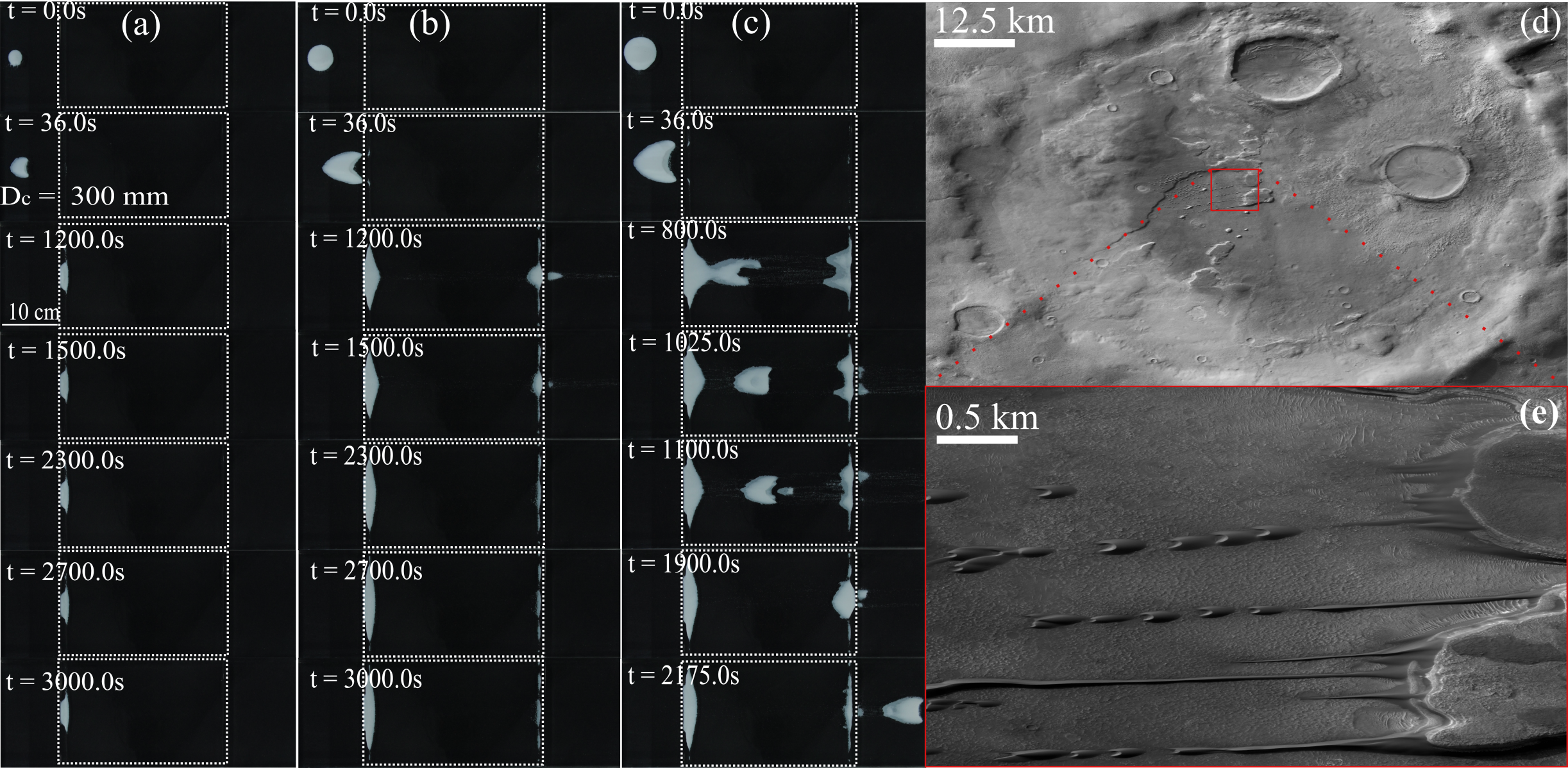}\\
	\end{center}
	\caption{Outcomes of interaction for large craters, $D/D_c$ $\leq$ 0.16. \corrb{(a)-(c) Snapshots stacked} vertically showing the time evolution for the Trapped, Trapped with mass loss, and Flow-induced fragmentation, respectively. The crater rim is highlighted by a dashed line in the snapshots, and the flow is from left to right. (d) HiRISE image (High Resolution Imaging Science Experiment, https://www.uahirise.org/[Dataset], n.d.) showing examples of Martian dunes interacting with large craters on the surface of Mars (-41.407° latitude, 44.587°E longitude, spacecraft altitude 253.9 km). Courtesy NASA/JPL-Caltech/UArizona. Panel (e) is a zoom on a region of panel (d), Courtesy NASA/JPL-Caltech/UArizona}
	\label{fig:snapshots_craters2}
\end{figure}

With the used setup, we could track the bedforms as they interacted with the three craters tested. Figures \ref{fig:snapshots_craters} and \ref{fig:snapshots_craters2} show snapshots placed side by side (Figures \ref{fig:snapshots_craters}) or stacked vertically (Figures \ref{fig:snapshots_craters2}) for some of our tests, for the smaller ($D_c$ $=$ 50 and 100 mm) and large ($D_c$ $=$ 300 mm) craters, respectively. The figures also show satellite images of Martian dunes in or around craters. Lighter versions of the movies corresponding to the snapshots of Figure \ref{fig:snapshots_craters} are available in the Supporting Information, and lighter versions of all movies used in the manuscript are available in an open repository \cite{Supplemental2}.

For all sizes of craters and bedforms tested, we typically observed five different outcomes:

\begin{itemize}
	\item \textit{Trapped}, which takes place when the bedform, once in the crater, remains blocked inside and halts its migration (most of the bedform grains remain trapped). Thus, this case is more frequent for small barchans (Figures \ref{fig:snapshots_craters}a and \ref{fig:snapshots_craters2}a).
	
	\item \textit{Trapped with mass loss}, which takes place when the bedform is trapped in the crater while part of its sediments are entrained further downstream (Figures \ref{fig:snapshots_craters}b and \ref{fig:snapshots_craters2}b). This case occurs for relatively small barchans, that is, larger than those in the Trapped case but still small enough to be trapped. In addition, this behavior becomes stronger as the fluid flow is increased.  
	
	\item \textit{Flow-induced fragmentation}, which occurs when the barchan breaks into smaller bedforms, with asymmetric mass redistribution and typically after and around the crater edges (Figures \ref{fig:snapshots_craters}c and \ref{fig:snapshots_craters2}c). This behavior is more frequent for moderate water velocities.
	
	\item \textit{Overpassing}, when the barchan goes over the crater and continues migrating further downstream, while part of its grains settle in the crater. At the end of the process, the barchan keeps its crescent shape (although with less grains).	This case typically occurs for large barchans (Figure \ref{fig:snapshots_craters}e).
	
	\item \textit{Overpassing with integrity loss}. In this case, the barchan goes over the crater but loses a considerable amount of its grains (which are entrained further downstream) when crossing the crater. At the end, the bedform is no longer a barchan, but a cluster of grains that disintegrates afterward (Figure \ref{fig:snapshots_craters}d).
\end{itemize}

\begin{figure}[h!]
	\begin{center}
		\includegraphics[width=.99\linewidth]{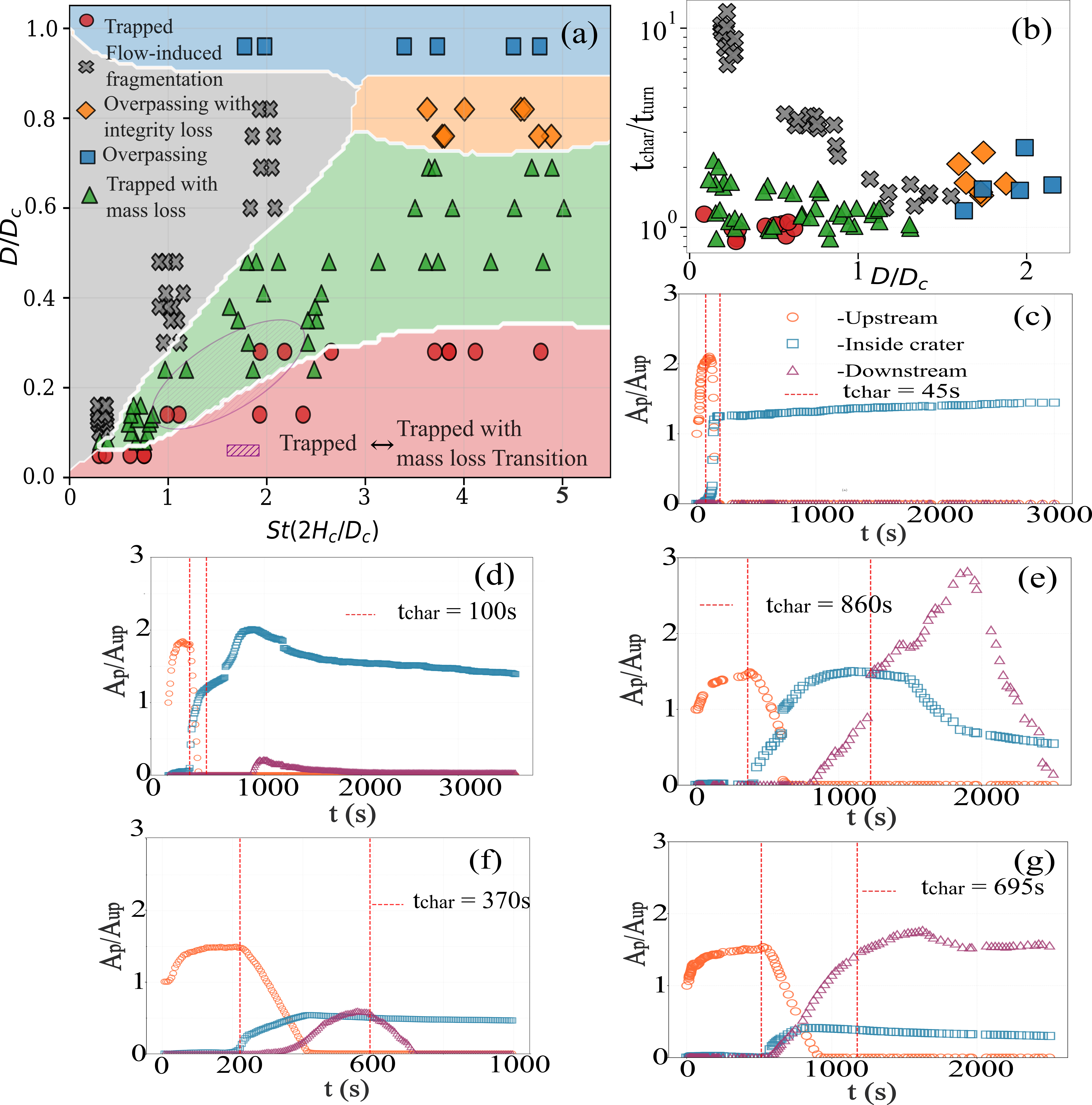}\\
	\end{center}
	\caption{(a)Different outcomes classified in the $D/D_c$ vs. St$(2H_c/D_c)$ space (barchan to crater size ratio and modified Stokes number, respectively). The symbols are listed in the key and the lines separating the different regions were drawn using SVM (support vector machine) and \textit{k}-nearest neighbors (KNN) for refinement. The hatched region corresponds to a transitional case between Trapped and Trapped with mass loss. \corrb{(b) Characteristic time} $t_{char}$ for completing an interaction with the crater, normalized by the turnover time $t_{turn}$, as a function of the $D/D_c$. The symbols are the same used in panel \textit{a}. (c)-(g) Time evolution of the projected area $A_p$ upstream, inside, and downstream the crater, shown for the crater–barchan interaction regimes: (c) Trapped, (d) Trapped with mass loss, (e) Flow-induced fragmentation, (f) Overpassing with integrity loss, and (g) Overpassing. The area $A_p$ (horizontal projection) is normalized by the initial upstream area $A_{up}$.}
	\label{fig:map_patterns}
\end{figure}

Based on the above characteristics for each observed pattern, we proceeded as in \citeA{Assis4} and inquired into dimensionless groups for classifying the patterns on an \textit{ad hoc} map. One group that arises naturally is the ratio between the barchan and crater equivalent diameters, $D/D_c$, since trapping occurs for small and overpassing for large barchans. The other parameter must take into account how close the grains follow the fluid flow (typically given by the Stokes number St based on the fluid velocity), which is in its turn disturbed by the crater. We propose, thus, a modified Stokes number given by the Stokes number St multiplied by the aspect ratio of the crater: St$(2H_c / D_c)$.

Figure \ref{fig:map_patterns}a shows the observed patterns classified in the $D/D_c$ vs. St$(2H_c/D_c)$ space, with the symbols listed in the figure key and different regions identified by making use of artificial intelligence (support vector machine - SVM, more information is available in the Supporting Information). We notice first that the patterns are grouped and well organized in the map, with the Trapped pattern occurring for small $D/D_c$ and the Overpassing pattern for large $D/D_c$. Then, Trapped with mass loss occurs for values of $D/D_c$ just above those for the Trapped case, but its region increases considerably with increasing the modified Stokes number until St$(2H_c / D_c)$ $\approx$ 3, from which value it seems to reach a plateau. Finally, between the Trapped with mass loss and Overpassing cases, Flow-induced fragmentation occurs for St$(2H_c / D_c)$ $\lesssim$ 3 and Overpassing with integrity loss for St$(2H_c / D_c)$ $\gtrsim$ 3, the latter having a region that seems independent of the modified Stokes number (while that for the former varies with it). We note that there is a lack of data on the bottom right of the classification map (small $D/D_c$ at large St$(2H_c / D_c)$) because bedforms are too unstable (they are swept by the water flow) at such flow conditions. On the top left (large $D/D_c$ at small St$(2H_c / D_c)$), the lack of data is due to the fact that the water flow is too weak for moving the larger barchans within the timescale of tests.

As shown in Figures \ref{fig:snapshots_craters}, \ref{fig:snapshots_craters2}, and \ref{fig:map_patterns}a, the observed patterns depend on both the relative size of barchans ($D/D_c$) and fluid flow. Thus, the mechanisms leading to those patterns can be understood by analyzing the fluid flow around and in the crater. Detailed measurements of the water flow inside a circular flat crater with a central peak were carried out by \citeA{Gundersen} using refractive index matching. As general features, they found that quasi-spanwise vortices are shed by the upstream rim, which impact the inner wall of the downstream rim. At the same time, two or four (depending on the distance from the crater floor) vertical counter-rotating vortices form inside the crater (symmetrically with respect to the centerline of the channel). Close to the crater center, those vortices direct horizontally the flow from the center toward the lateral walls, and from there to the upstream (in the upstream half of the crater) and downstream rims (in the downstream half). Therefore, in the circular craters of our tests, barchans whose volume do not occupy approximately half of the crater (when spread inside) remain trapped. Those that occupy more than half of the crater area, but do not fill the entire crater, are susceptible to both the vertical vortices existing in the downstream half and those impinging the downstream rim: the bedforms then break into smaller bedforms that are entrained further downstream. Finally, when barchans occupy the entire crater volume, the grains that exceed the crater are entrained further downstream. In the case of the larger crater ($D/D_c$ $\leq$ 0.03) the flow corresponds to a typical flow downstream a step, for which there exists a recirculation region of length of approximately 6 to 10 times the step height (between 30 and 50 mm, in our case) just downstream the step \cite{Goldstein, Etheridge}. The dunes for which all grains remain in the region upstream the reattachment point remain, thus, trapped, while those that are larger and have part of their grains downstream the reattachment point are partially entrained further downstream (those grains downstream the reattachment point). That is why the quantity of grains that remain in the crater at the end of the process is the same in the cases of Figures \ref{fig:snapshots_craters2}b and \ref{fig:snapshots_craters2}c, although the initial sizes are different.

Having classified the observed outcomes, we now inquire into the spreading/shrinkage of the bedform and the characteristic times of interaction. For the spreading/shrinkage of bedforms, we do not have access to their three-dimensional structure, since we acquired top-view images only. For that reason, we decided to use the surface area of bedforms projected on the horizontal plane, $A_p$, as an indicator of how spread the grains are on that plane. Figures \ref{fig:map_patterns}c-g show the time evolution of $A_p$ for the Trapped (Figure \ref{fig:map_patterns}c), Trapped with mass loss (Figure \ref{fig:map_patterns}d), Flow-induced fragmentation (Figure \ref{fig:map_patterns}e), Overpassing with mass loss (Figure \ref{fig:map_patterns}f), and Overpassing (Figure \ref{fig:map_patterns}g) cases. In these figures, we plotted the areas in three different regions as functions of time: upstream, inside, and downstream the crater (the symbols are listed in the figure key). For the Trapped case, Figure \ref{fig:map_patterns}c shows that the projected area in the upstream region initially increases due to the deformation of the conical pile into a barchan dune, and then decreases to zero as the barchan enters the crater. While the upstream values are decreasing, the area in the crater increases until reaching a plateau, and the area downstream remains equal to zero at all times, showing that the bedform is completely blocked inside the crater. For the case with mass loss, Figure \ref{fig:map_patterns}d shows a similar behavior, the only difference being that the area downstream increases and then vanishes as material leaves the crater. On the other extreme, Figure \ref{fig:map_patterns}g shows an initial increase of the upstream area (due to deformation of the initial pile into a barchan shape) and then a decrease as the barchan goes over the crater. While the upstream area decreases, those inside the crater and downstream of it increase, showing that the barchan is overpassing the crater. At the end, both those inside the crater and downstream of it reach a plateau, showing that a small part of grains remains trapped in the crater while the barchan continues its migration further downstream. For the case with mass loss, Figure \ref{fig:map_patterns}f shows a similar behavior, the only difference being a faster decrease in the downstream area due to material loss (which in this case decreases to zero as the bedform leaves the field of view). For the Flow-induced fragmentation, Figure \ref{fig:map_patterns}e shows that the upstream area increases and decreases (as for the other cases, due to the deformation of the initial pile into a barchan shape) until disappearing. While it decreases, the area inside the crater increases, and then reaches a constant value for a certain time interval in which the downstream area begins increasing. Afterward, the area inside the crater decreases while the downstream area continues increasing until reaching a value almost three times that of the initial barchan, reflecting a spreading of bedforms into diverse smaller forms. Finally, the downstream area decreases toward zero as the small bedforms leave the field of view of the camera, and the area inside the crater tends to reach a plateau because of grains that remain trapped in the crater.

We also computed a characteristic time $t_{char}$ for the observed patterns to completely cross or enter the crater, corresponding to the duration of interaction. For that, we defined the duration as the time interval between the first time when the downstream region of the barchan enters the crater (the horns or the lee side, depending on the case) and that when either the bedform toe leaves the crater (crossing cases) or the bedform remains completely trapped (trapped cases). We then normalized the measured times by the turnover time, $t_{turn}$ $=$ $L/C$, where $C$ is the bedform celerity and $L$ its length (distance between the toe and the lee face, along the centerline) when upstream the crater. An example of snapshots showing the identification of initial and final times is available in Figure S5 of Supporting Information. Figure \ref{fig:map_patterns}b shows the characteristic time in dimensionless form, $t_{char}/t_{turn}$, as a function of the barchan-crater aspect ratio $D/D_c$. We observe that, excepting for Flow-induced fragmentation, the characteristic time increases with increasing the relative barchan size, from  roughly $t_{char}/t_{turn}$ $\approx$ 1 when $D/D_c$ $\lesssim$ 1 to roughly $t_{char}/t_{turn}$ $\approx$ 2 when $D/D_c$ $\gtrsim$ 1 (although with considerable scatter). In this trend, as $t_{char}/t_{turn}$ increases with increasing $D/D_c$, the interaction patterns change (since they depend on the size ratio). We observe an inverse trend for Flow-induced fragmentation, the characteristic time decreasing with $D/D_c$ from much larger values when $D/D_c$ $\lesssim$ 1 ($t_{char}/t_{turn}$ $\approx$ 10 when $D/D_c$ $\rightarrow$ 0). This is a consequence of the larger distance bedforms have to migrate within the crater for larger craters (the time becoming larger for relatively large craters). In all cases, $t_{char}/t_{turn}$ $\rightarrow$ $\approx$ 2 as $D/D_c$ $\rightarrow$ 2, indicating that when barchans are larger than the crater they need to complete two turnovers to overpass the crater (pattern that arises for larger values of $D/D_c$).

\section{Extrapolation of subaqueous results to Martian dunes}

Finally, we note the similarities and differences between the tested conditions and Martian environment. The first observation is that subaqueous ripples and dunes are shaped by the direct action of the fluid flow, such as happens for eolian dunes. In our experiments, the water flow was in the hydraulic-smooth regime, which was recently shown to be likely the same regime found in Martian environment \cite{Alvarez7}. This means that in both the subaqueous and Martian environments the pertinent length of the fluid flow close to the bed surface is the viscous length $\nu/u_*$. Therefore, the fluid flow shaping the subaqueous bedforms in our experiments are comparable, to some extent, to Martian dunes. The second observation concerns the crater shapes used in our study. Craters on Mars and on Earth usually have an elevated rim, which disturbs even more the fluid flow \cite{Greeley}. In addition, small craters have a bowl shape (known as simple craters), while larger craters have a flat bottom, sometimes with a central peak or peak rings \cite{Arvidson, Melosh, Suarez, Barlow}. In this work, we are particularly interested on the effects caused by the cavity and decided to simplify the crater geometry (we note that this is the first time that such dune-crater interactions are studied in laboratory). Therefore, we used a flat-bottomed shape without central peaks, the effects caused by the elevated rim and central peak being left for a future work. However, the size of craters used in our experiments are relatively small with respect to the size of barchans, with 0.09 $\leq$ $D/D_c$ $\leq$ 0.96 while on Mars we find $D/D_c$ $\lesssim$ 0.1 for flat-bottomed craters \cite{Robbins}. This means that we should have used larger flat-bottomed craters, or deeper bowl craters (for the tested $D/D_c$ values, real craters have a bowl shape with $H_c/D_c$ $\sim$ 0.2, while in our experiments 0.03 $\leq$ $H_c/D_c$ $\leq$ 0.10). Thus, extrapolations of the present results to dune-crater interactions on Mars must take into account that disturbances of the fluid flow are expected to be stronger in the Martian case.

A last observation is that our results are valid for subaqueous bedforms, for which the density ratio between sand and water $\rho_s/\rho$ is of order one, so that sand particles are entrained by rolling, sliding, or effectuating small jumps, while on Mars $\rho_s/\rho$ is of the order of 10$^5$ and grains are entrained by saltation. Under those conditions, the saturation length $L_{sat}$ (length over which a varying fluid flow takes to be in equilibrium with sand entrainment, and, thus, erosion is expected) is of approximately 1-10 mm under water and of the order of 1 m on Mars \cite{Duran4}. This corresponds to both $L_{sat}/D_c$ and $L_{sat}/D$ within 10$^{-2}$ and 10$^{-1}$ in our experiments, and both $L_{sat}/D_c$ and $L_{sat}/D$ $\lesssim$ 10$^{-3}$ on Mars, meaning that bedform erosion tends to be higher in our experiments. Although hints for Martian dunes can be obtained from our findings, extrapolations when $D/D_c$ $\sim$ 1 must be carried out with caution.

\section{Conclusions}

We found that a subaqueous barchan moving over a crater-like depression can have one of the following outcomes: (i) be completely blocked (Trapped) when the barchan is much smaller than the crater dimensions; (ii) be trapped in the crater while part of its sediments is entrained further downstream (Trapped with mass loss). This case occurs for barchans larger than those in the trapped case, but small enough to remain trapped; (iii) cross the crater (Overpassing), when the barchan is equal to or larger than the crater dimensions. The barchan goes over the crater and continues migrating further downstream, while part of its grains settle in the crater; (iv) cross the crater but losing a considerable amount of grains and the crescent shape (Overpassing with integrity loss). This case occurs for barchan sizes within those for the trapped with mass loss and overpassing, when the fluid disturbances are relatively strong; (v) or fragment into smaller bedforms (Flow-induced fragmentation), with asymmetric mass redistribution and occurring typically after and around the crater edges. This case also occurs for barchan sizes within those for the Trapped with mass loss and Overpassing, but when the fluid disturbances are relatively weaker. Based on these observations, we propose a map in the barchan-crater size ratio ($D/D_c$) and modified Stokes number (St$(2H_c / D_c)$) space that classifies the different outcomes. We also found that, with the exception of Flow-induced fragmentation, the characteristic time $t_{char}$ for the barchan-crater interaction increases from 1 to roughly 2 times the turnover time $t_{turn}$ as $D/D_c$ increases. For Fragmentation with mass loss, $t_{char}$ decreases from approximately 10 to 2 times $t_{turn}$ as $D/D_c$ increases (varying, thus, by one order of magnitude). Our findings shed new light on the behavior of Martian dunes interacting with craters; however, any extrapolation must be done with caution, since on Mars the density ratio between the grains and fluid is much higher.

\section*{Conflict of Interest}
	The authors declare no conflicts of interest relevant to this study.

\section*{Data Availability Statement}
\begin{sloppypar}
	Data (movies) supporting this work were generated by ourselves and are available in Mendeley Data \cite{Supplemental2} under the CC-BY-4.0 license.
\end{sloppypar}

\acknowledgments
\begin{sloppypar}
The authors are grateful to FAPESP (Grant Nos. 2018/14981-7 and 2022/01758-3) and to CNPq (Grant No. 405512/2022-8) for the financial support provided. The authors thank Alcimar da Silveira for his technical support. 
\end{sloppypar}

\nocite{fix1985discriminatory, cover1967nearest, Cover1965, rhys2020machine, pedregosa2011scikit}
\bibliography{references}

\end{document}